\documentclass[sigconf]{acmart}
\def\BibTeX{{\rm B\kern-.05em{\sc i\kern-.025em b}\kern-.08emT\kern-.1667em\lower.7ex\hbox{E}\kern-.125emX}}

\usepackage[compact]{titlesec}
\usepackage[utf8]{inputenc}
\usepackage{enumitem, enumerate}
\usepackage{blindtext}
\usepackage{tikz}
\usepackage{color}
\usepackage{colortbl}
\usepackage{textgreek}
\usepackage{algorithm}
\usepackage[T1]{fontenc}
\usepackage{algorithmicx}
\usepackage{algpseudocode}
\usepackage{subcaption}

\newcommand{\name}{\textsc{PoisonIvy}\xspace}

\definecolor{Gray}{gray}{0.9}
\definecolor{LightCyan}{rgb}{0.88,1,1}
\newcommand*\circled[1]{\tikz[baseline=(char.base)]{
            \node[shape=circle,fill,inner sep=1pt] (char) {\textcolor{white}{#1}};}}

\copyrightyear{2020}
\acmYear{2020}
\setcopyright{acmcopyright}
\acmConference[BuildSys '20]{The 7th ACM International Conference on Systems for Energy-Efficient Buildings, Cities, and Transportation}{November 18--20, 2020}{Virtual Event, Japan}
\acmBooktitle{The 7th ACM International Conference on Systems for Energy-Efficient Buildings, Cities, and Transportation (BuildSys '20), November 18--20, 2020, Virtual Event, Japan}
\acmPrice{15.00}
\acmDOI{10.1145/3408308.3427606}
\acmISBN{978-1-4503-8061-4/20/11}

\begin{document}

\title[\name: (In)secure Practices of EIoT Systems in Smart Buildings]{\name: (In)secure Practices of Enterprise IoT Systems\\ in Smart Buildings}

\author[L. Puche Rondon, L. Babun, A. Aris, K Akkaya, A. Uluagac]{Luis Puche Rondon, Leonardo Babun, Ahmet Aris, Kemal Akkaya, and A. Selcuk Uluagac}
\affiliation{%
  \institution{Cyber-Physical Systems Security Lab\\
  Department of Electrical and Computer Engineering\\
  Florida International University, Miami, Florida, USA
  }
}

\email{Email: {lpuch002, lbabu002, aaris, kakkaya, suluagac }@fiu.edu}


% --------------------- ABSTRACT ----------------------------

\begin{abstract}

The rise of IoT devices has led to the proliferation of smart buildings, offices, and homes worldwide. Although commodity IoT devices are employed by ordinary end-users, complex environments such as smart buildings, government, or private smart offices, conference rooms, or hospitality require customized and highly reliable solutions. Those systems called Enterprise Internet of Things (EIoT) connect such environments to the Internet and are professionally managed solutions usually offered by dedicated vendors (e.g., Control4, Crestron, Lutron, etc.). As EIoT systems require specialized training, software, and equipment to deploy, many of these systems are closed-source and proprietary in nature. This has led to very little research investigating the security of EIoT systems and their components. In effect, EIoT systems in smart settings such as smart buildings present an unprecedented and unexplored threat vector for an attacker. In this work, we explore EIoT system vulnerabilities and insecure development practices. Specifically, focus on the usage of drivers as an attack mechanism, and introduce PoisonIvy, a number of novel attacks that demonstrate how it is possible for an attacker to easily attack and command EIoT system controllers using malicious drivers. Specifically, we show how drivers used to integrate third-party services and devices to EIoT systems can be trivially misused in a systematic fashion. To demonstrate the capabilities of attackers, we implement and evaluate PoisonIvy using a testbed of real EIoT devices in a smart building setting. We show that an attacker can easily perform DoS attacks, gain remote control, and maliciously abuse system resources (e.g., bitcoin mining) of EIoT systems. Further, we discuss the (in)securities in drivers and possible countermeasures. To the best of our knowledge, this is the first work to analyze the (in)securities of EIoT deployment practices and demonstrate the associated vulnerabilities in this ecosystem. With this work, we raise awareness on the (in)secure development practices used for EIoT systems, the consequences of which can largely impact the security, privacy, safety, reliability, and performance of millions, if not billions, of EIoT systems worldwide.  
\end{abstract}

\maketitle

% --------------------- INTRODUCTION ----------------------------

\section{Introduction}

The introduction of modern commodity IoT devices has changed the everyday lives of users with the deployment of millions of smart environments (e.g., smart buildings, offices, homes, etc.) worldwide \cite{SmartHomesUSEurope}. While many IoT systems are easily installed by average end-users via Do-it-Yourself (DIY) applications, Enterprise Internet-of-Things (EIoT) systems exist as an automation solution for professional settings. As such, EIoT systems are used exclusively for applications such as smart buildings, luxury smart homes, expensive yachts, classrooms, meeting rooms, government offices, and business establishments. In these professional settings, proprietary EIoT systems (e.g., Crestron, Control4, and Savant) introduce a robust, reliable, and custom solutions catered to meet an enterprise client's needs. As such, EIoT systems require professional installation and specialized training to deploy. Additionally, maintenance, upgrades, and service of EIoT systems is handled by specialized integrators and not the end-users.

Although many consumer-grade commodity IoT systems are well-understood due to their mainstream popularity, very little security research exists on EIoT systems' design, development, verification processes, and vulnerabilities. The lack of research on these systems has led many users to overlook EIoT systems as possible attack vectors and assume that these systems are secure. With many of these professional systems present in high-profile locations, evaluating threats for EIoT systems should be of utmost importance. In this paper, we systematically explore EIoT system vulnerabilities and insecure development practices, specifically, the usage of drivers as an attack mechanism. In order to demonstrate that malicious actors can easily attack EIoT systems, we introduce \name, a collection of novel attacks that leverages EIoT system vulnerabilities to an attacker's benefit. Specifically, we attack one of the integral components of EIoT systems: drivers, which contain all of the necessary software to integrate external software and devices into EIoT ecosystems. To show that EIoT systems may be attacked through drivers, we analyze the highly-programmable nature of drivers and the associated vulnerabilities. With \name, we show that it is feasible to use malicious code in drivers to perform attacks using EIoT systems. As many third-party devices do not have verified drivers, installers must sometimes opt for unverified drivers with no method to guarantee their safety, making \name a real and viable threat against EIoT systems. 

To raise awareness on the (in)secure development of the drivers that control EIoT systems, and how vulnerabilities may impact the security of smart buildings and other relevant settings, we perform \name attacks in a realistic EIoT system testbed in a smart building setting. For this, we show how with \name an attacker can use EIoT system drivers to assume arbitrary control of device functions in smart buildings remotely. Specifically, with \name, an attacker may remotely (1) perform Denial-of-Service (DoS) attacks on EIoT system, (2) assume control EIoT systems as an effective botnet, and (3) use EIoT system resources to perform illicit activities (e.g., bitcoin mining, distributed password cracking, etc.). As drivers are outside of any traditional protection mechanisms, there are no defense mechanisms against attacks in EIoT systems. To the best of our knowledge, this is the first work to analyze the (in)security of the EIoT deployments and clearly demonstrate the vulnerabilities in this ecosystem which can result in consequences that can largely impact the security of EIoT systems deployed in millions of smart buildings. 

\noindent\textbf{Contributions: } The contributions of this work are as follows:
\begin{itemize}
    \item We demonstrate that EIoT system drivers are a viable attack vector for smart buildings by introducing \name, a series of novel attacks against EIoT systems. 
    \item We test and evaluate \name attacks in a real EIoT system and leverage malicious drivers to cause undesired behavior in a smart building on behalf of a remote attacker.
    \item We articulate the effects and implications of insecure EIoT systems, their secure development, verification, and we open the discussion to the best practices and potential countermeasures to \name attacks. 
\end{itemize}

\noindent\textbf{Organization:}
The rest of this work is organized as follows: In Section \ref{sec:Background}, we provide background information on EIoT systems. Section \ref{sec:ThreatModel} introduces the problem scope and threat model of the \name attacks. The design and implementation of \name attacks are introduced in Section \ref{sec:Architecture}. The testbed design, network configuration, software, and \name attack implementation are covered in Section \ref{sec:Implementation}. Additionally, this section highlights the findings and implications of \name-based attacks. In Section \ref{sec:Discussion}, we discuss possible mitigation strategies and defense mechanisms against \name. Section \ref{sec:RelatedWork} discusses the related work. Finally, we conclude the paper in Section \ref{sec:Conclusions}.

% --------------------- BACKGROUND ----------------------------

\section{Background}
\label{sec:Background}

In this section, we introduce some necessary background concepts about smart buildings and EIoT systems relevant to \name.

\subsection{EIoT Systems}

\begin{figure}[t]
\centering{\includegraphics[width=0.45 \textwidth]{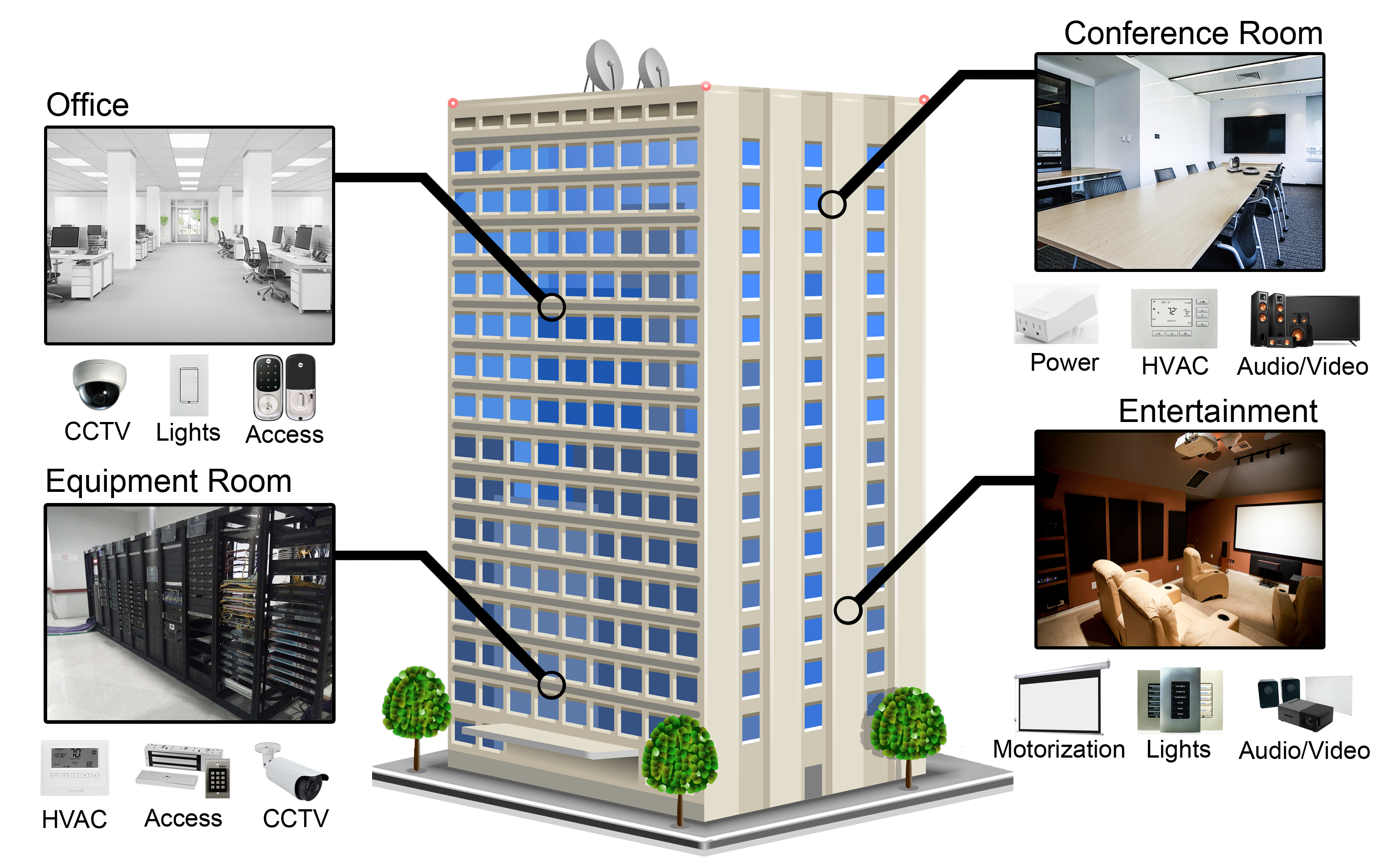}}
    \caption{Applications of EIoT systems in smart buildings.}
\label{fig:smart-building}
\vspace{-0.25in}
\end{figure}

The need for automation in smart buildings, luxury homes, commercial, and industrial applications has existed since the 70's \cite{SmartTS}. As such, there are many different use-cases where EIoT is the best solution for automation and integration of multiple devices. Automation may be done in a single-room systems (e.g., a theater, a conference room) or have multiple rooms or floors under the same system. The expandable nature of EIoT systems allows for small or large smart systems and integration between these systems. Figure \ref{fig:smart-building} highlights applications of EIoT in smart buildings. For instance, a smart office may be automated with CCTV systems, lighting control systems, and access control components with an EIoT system. As such, EIoT systems are customized for each specific application and deployment. We highlight some EIoT use-cases on smart buildings, where these use cases can work together under a single EIoT system. As such, if the EIoT system is compromised, the integrated devices may also be compromised.

\noindent\textbf{Lighting Control.} Any control of physical lighting or electrical loads by an EIoT system (e.g., lights, fans, outlets). EIoT systems may be used in this use-case to schedule light events, program independent keypads, and allow remote control of lighting functions. EIoT allows users to control their lights remotely, schedule light events (e.g., wake up, turn outdoor lights on sundown, and trigger light-based events from other devices.

\noindent\textbf{Security and Safety.} EIoT systems are often integrated to control security components. This integration grants authorized users the ability to control security aspects of a location (e.g., CCTV systems, access control systems, motion sensors, fire alarms, security alarm systems). As such, EIoT systems allow for remote access, control, and camera activation based on motion sensor triggers. EIoT allows users to integrate other components such as lights with security systems. For example, an EIoT system may start flashing lights when the alarm system is triggered.

\noindent\textbf{Advanced Media Control.} The control and management of media and audio/video (A/V) components with EIoT systems (e.g., projectors, televisions, video distributions, HDMI networks, audio matrix management). EIoT systems manage complex audio/video distribution networks from a single interface through audio/video zones, audio switchers, video switchers, and amplifiers. With the complexity of many A/V systems, EIoT is a reliable method of control through a single user interface.

\subsection{Architecture of EIoT Systems}

\begin{figure}[t]
\centering{\includegraphics[width=0.35 \textwidth]{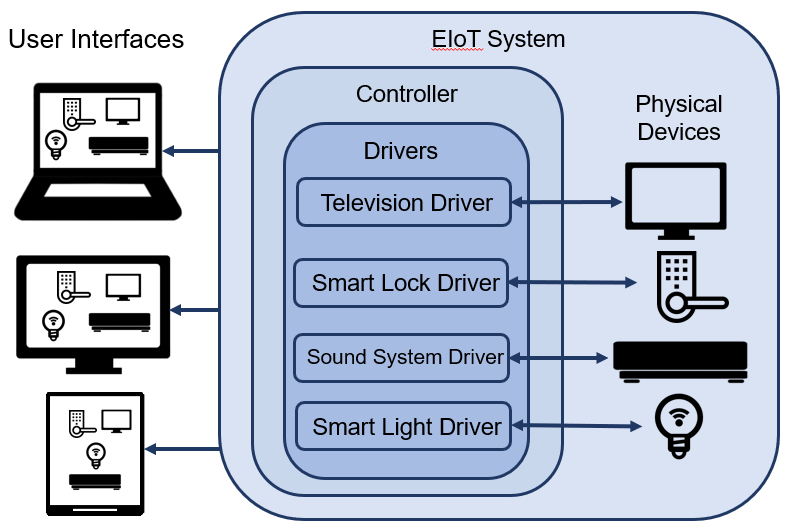}}
    \caption{EIoT system with four different control drivers, controller, and user interfaces. Individual devices are controlled through the user interfaces after being integrated.}
\label{fig:smart-system}
\vspace{-0.2in}
\end{figure}

As all installations are custom, \textit{an integrator}, a specialized programmer and installer, configures and integrates all devices with an EIoT system. An integrator is hired to perform the physical installation, device configuration, testing, and technical support of the EIoT system \cite{crestrondeploy, control4deploy}. EIoT systems follow a centralized system design. We refer to Figure \ref{fig:smart-system} which shows the generalized design of an EIoT system. These systems consist of several modules, the user interfaces, a controller, drivers, and physical devices. \textit{Physical devices} are devices such as televisions or smart lights which are integrated into an EIoT system. EIoT systems usually integrate physical devices without cloud-based access \cite{control4deploy}. To accomplish this, EIoT systems use a central \textit{controller}, a dedicated processing device that contains the main logic, communication behavior, and configuration of an EIoT system. The controller stores \textit{drivers} locally, which contain all the necessary information to integrate a specific physical device or service into the EIoT system. Each physical device requires a driver of its own to be integrated to a smart system. For instance, to integrate a smart door lock, an EIoT driver for that specific door lock must be added to the controller. After a device is integrated, that device becomes available to the user through any of the smart system's user interfaces (e.g., laptop app, phone/tablet app, television on-screen-display, etc.). In contrast to traditional system drivers, EIoT drivers function on top of the proprietary EIoT operating system. The implementation may also be different between EIoT systems.

\subsection{EIoT Drivers} 
\label{subsec:EIoTDrivers}

One of the primary components of many EIoT systems is the inclusion of \textit{drivers} which may have different names depending on the manufacturer (e.g., Crestron modules, Control4 drivers). Drivers provide all the information and software modules necessary to integrate a device into a centralized EIoT system. For instance, to integrate a Sony television into an EIoT system, the controller must know what the protocol of communication is (e.g., IR, Serial, IP, ZigBee, Zwave), the physical inputs (HDMI ports, analog ports, etc), and the vendor-specific proprietary commands to interface with a device. Drivers are not limited to simply integrating physical devices and they also integrate services such as Weatherbug to add more functionality to an existing system \cite{control4composerrelease, control4driversearch}. 

\textbf{Vendor Drivers.} Drivers are inserted and configured during programming or maintenance stages of a smart environment by the integrator. Integrators may obtain drivers in three different ways: (1) they may get drivers directly from the EIoT system software (pre-loaded drivers), (2) directly from a catalog hosted by the manufacturer of the EIoT system devices, or (3) download from a third-party site in the Internet (from a third-party vendor or a developer). Vendors of EIoT systems often validate drivers distributed in their platforms for functionality such as Control4's certified drivers \cite{control4certified}. However, with millions of different devices to be integrated, certifying every driver is not possible. \textit{In effect, integrators may be forced to use third-party drivers for their installations if no drivers are available for their specific solutions from the vendor or manufacturer. In this work, we focus on unverified drivers, or drivers available on third-party sites that have not been checked for malicious content.}

\textbf{Driver Verification Mechanisms.} Many operating systems and platforms offer signature and verification mechanisms to guarantee the authenticity and integrity of software components. Microsoft uses digital signatures to guarantee the integrity of Microsoft drivers \cite{MicrosoftSigning}. Apple requires XCode and developer ID certificates to sign software available for MAC computers \cite{AppleDevId}. In Linux, the kernel module signing facility secures Linux modules with signatures before installation \cite{LinuxSigning}. Further, Android developers have the ability to sign apps to guarantee the integrity of installed applications \cite{AndroidSigning}. \textit{In contrast to these well-documented practices, EIoT vendors do not offer validation for EIoT drivers for their systems. As EIoT drivers often operate strictly on the proprietary software, traditional hardware-level driver defenses do not apply to application-layer based EIoT drivers. As such, integrators are forced to trust unverified software which may be malicious in nature.}

\textbf{Unverified Drivers. } Integrators may opt for unverified, third-party drivers due to several reasons:
\begin{itemize}
    \item\textit{Driver Availability. } Verified drivers may not always be available to an integrator. Therefore, the only recourse to integrate a third-party device to an EIoT system may be with an unverified driver from an untrusted source. Additionally, to integrate less-known devices, the driver has to be made by the manufacturer, who may be untrusted and their code closed-source. For instance, available integrator forums, offer a floury of unverified drivers for projectors, televisions, and other devices \cite{control4forums}. Additionally, many vendors do not offer EIoT drivers for their devices, leading to third-party developers to offer their own drivers.
    \item\textit{Cost. } Developers may charge for verified drivers (e.g., Atlona HDMI Switcher drivers for 110 USD), which in turn has to be paid by the integrator and end-user \cite{DriverCentral}. Integrators may be tempted to use free unverified drivers available on forums and online storefronts. Further, while paid drivers may be made by trusted developers, they are not necessarily verified by the EIoT vendor.
    \item\textit{Compatibility. } Devices may change commands and specifications when their firmware is updated \cite{appledriver}. As such, verified drivers need to be updated to remain compatible with the latest models and firmware. To get a system running quickly after an update, an integrator may use an unverified driver that claims to run perfectly with a newer firmware version of the device when a verified driver is not available. 
    \item\textit{Phishing. } It is possible that an integrator may install an untrusted driver through a phishing link offering a ``driver update'' or a tampered vendor website. It is possible to receive drivers through email attachments, impersonating a trusted vendor.
\end{itemize}

% --------------------- THREAT MODEL ----------------------------

\vspace{-0.1in}
\section{Problem Scope and Threat Model}\label{sec:ThreatModel}

\newcommand{\threatdos}{Denial-of-Service\xspace}
\newcommand{\threatbotnet}{Remote Control\xspace}
\newcommand{\threatfarming}{Malicious System Resource Farming\xspace}

This section presents the problem scope and threat model for \name attacks. 

\subsection{Problem Scope}

\begin{figure*}[!th]
\centering{\includegraphics[width=0.75\textwidth]{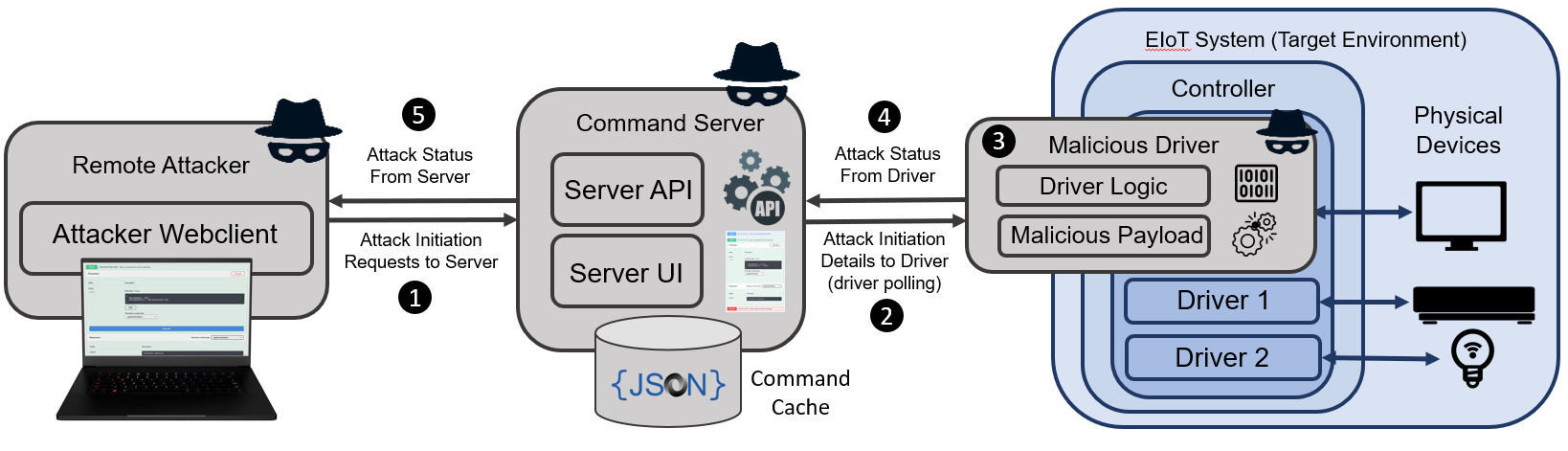}}
\caption{General end-to-end implementation for \name-based attacks. Attack-related components are highlighted in gray, EIoT system components are in blue.}
\label{fig:architecture}
\vspace{-0.15 in}
\end{figure*}

This work assumes the existence of an EIoT system installed in a smart building. Indeed, such EIoT systems have experienced a rapid increase in popularity in smart buildings, luxury smart homes, expensive yachts, classrooms, meeting rooms, government offices, and business establishments. The EIoT system's controller is connected to a network and the Internet. The attacker, named Mallory, is a malicious actor with knowledge of EIoT systems and their weaknesses. In this scenario, Mallory develops a malicious driver for a popular device and advertises the driver through user boards such as online forums \cite{control4forums} to integrators. Mallory also creates fake accounts to give good reviews on the driver and mislead integrators. The driver advertised is not available by the manufacturer or through verified drivers, making Mallory's driver the only way to integrate a particular device into an EIoT system. With this malicious driver, Mallory assumes the control of EIoT system controllers and uses her machine to execute remote attacks.

Additionally, we assume that an integrator uses Mallory's unverified malicious driver for the EIoT system deployment, introducing it into the system without issues as there are no security mechanisms in place. These assumptions are realistic as online drivers from third-party sites are not verified, and smart systems require Internet connectivity for many of their services (e.g., remote access, music streaming, movies, etc.) \cite{control4internet}. Anyone can upload a driver to public forums easily. As integrators may download unverified drivers from any website, an attacker can create an attractive driver for integrators to download and install in their systems. For instance, unverified drivers may be offered at a third-party website which advertises them. In our current scenario, Mallory compromises EIoT system devices indirectly through the use of a downloaded unverified malicious driver.

The consequences of driver-based attacks depend on the capabilities of drivers in a specific EIoT system. Drivers with network capabilities may be used to attack servers and other devices when multiple controllers are infected by the driver. Additionally, access to system resources would allow Mallory to perform cryptographic operations in infected controllers. In effect, Mallory could use infected devices to mine cryptocurrency or perform other cryptographic-based operations (e.g., password cracking) \cite{botnetmining}. Moreover, Mallory could simply use a driver in an attempt to overwhelm the host system, causing a local DoS condition. Any devices integrated into the EIoT system may become unreachable through user interfaces. It is also notable that drivers may act as a ``bridge'' between traditional IP networks and other protocols. A driver may have the capability to communicate with devices with embedded protocols (e.g., HDMI's Consumer Electronics Control protocol (CEC), Serial, InfraRed (IR)) making previously unreachable devices reachable. For instance, work on the topic of HDMI-CEC has demonstrated that arbitrary CEC control makes attacks on multiple connected HDMI devices viable \cite{hdmiwalk}. 

\subsection{Threat Model}

In our work, we consider the following powerful adversaries as part of the threat model.

\textit{Threat 1: \threatdos. } This threat considers DoS attacks where Mallory disrupts the availability of an EIoT system through the use of a malicious driver.

\textit{Threat 2: \threatbotnet. } This threat considers a case where Mallory assumes the control of EIoT system devices to execute DoS attacks on local/remote devices or webservers.

\textit{Threat 3: \threatfarming. } In this threat, Mallory uses local system resources in a compromised device to perform unauthorized processor-intensive actions to her benefit (e.g., cryptocurrency mining \cite{botnetmining}, password cracking \cite{botnetscracking}) 

Note that this work does not consider attacks that focus on traditional Linux or other mainstream operating system malware. Similarly, this work does not consider protocol-based vulnerabilities (e.g., Zigbee vulnerabilities). 

% --------------------- ARCHITECTURE ----------------------------

\section{{\name Architecture}}
\label{sec:Architecture}

\newcommand{\moduleonetitle}{Remote Attacker\xspace}
\newcommand{\moduletwotitle}{Command Server\xspace}
\newcommand{\modulethreetitle}{Malicious Driver\xspace}
\newcommand{\modulefourtitle}{Target Environment\xspace}

\newcommand{\moduleone}{remote attacker\xspace}
\newcommand{\moduletwo}{command server\xspace}
\newcommand{\modulethree}{malicious driver\xspace}
\newcommand{\modulefour}{target environment\xspace}

To demonstrate EIoT drivers as a viable threat vector, we developed \name, a series of driver-based attacks. In this section, we detail the end-to-end implementation of \name attacks, which become practical and applicable due to EIoT component implementations not built with security in mind. Such implementation involves the interaction of four modules: \textit{\moduleone, \moduletwo, \modulethree, and the \modulefour}. 

\subsection{\name Overview}

The proposed end-to-end implementation of \name is highlighted in Figure \ref{fig:architecture}. In this architecture, the integrator has unknowingly installed the malicious unverified driver due to the reasons outlined in Section \ref{subsec:EIoTDrivers}, and the EIoT system controller has been compromised. As explained earlier, this could be achieved through a forum post advertising a malicious driver as benign. The attack begins with a \moduleone (e.g., Mallory) initiating an attack with a webclient such as a laptop by communicating with the  \moduletwo \circled{1}. The \moduletwo grants Mallory an intermediary point of communication between her device and infected controllers, and includes three components: the \textit{server API}, \textit{server UI}, and the \textit{command cache}. The server API represents the primary endpoints (e.g., REST Architecture) of the server, which can be requested by Mallory or the malicious driver. Mallory uses the server UI component executed by the webclient, which grants her a visual interface, to initiate attacks and view the attack's status. As the last component of the \moduletwo, the command cache stores attack initiation requests fetched by malicious drivers through the server API endpoints. 

Once the command server receives initiation requests from Mallory, the malicious driver can now query the command server for new attack details \circled{2}. The Malicious driver is the core of \name attacks and contains the \textit{driver logic} and the \textit{malicious payload}.  As such, the driver logic controls a smart device in an expected manner, which allows the driver to appear as a benign driver. In contrast, the malicious payload contains the attacker's malicious code for the execution of the attacks \circled{3}. Finally, the target environment contains the smart system's controller and the drivers of the smart environment. Mallory takes control of the target environment's functions through the malicious driver. As attacks complete, the malicious driver sends back the attack status to the \moduletwo \circled{4}. The attack status includes feedback to the attacker from the driver, such as hashing results, errors, or success codes, and can be then queried by Mallory from the \moduletwo \circled{5}.

\textbf{\moduleonetitle. } In \name, the \moduleone is the malicious actor of the attacks. The primary purpose of the \moduleone is to send commands to the \moduletwo to be executed by a driver-compromised controller. In this case, Mallory uses the attacker webclient, which is any web-enabled device such as a laptop, tablet, or phone that is used to initiate the attack. Additionally, the remote attacker receives information from the \moduletwo such as attack results, hashing status, or available controllers to use for attacks.

\textbf{\moduletwotitle.} The \moduletwo module acts as an intermediary communication point between \moduleone and \modulethree. Controlled devices query the \moduletwo for new attacks to execute.  Additionally, the \moduletwo is divided into three components: the \textit{server API logic}, the \textit{server UI}, and the \textit{command cache}. The server API component contains all the API programming logic and REST paths needed for communication between the \moduleone and the \modulethree. The server UI component, in contrast, is an interface for Mallory to interact with and view reports of devices. These reports may contain results on successful hashing attempts, controlled device status, or the current status of an attack. The server UI component, depending on the type of attack performed, could be implemented as a fully-dynamic website (to view reports on attacks) or as a simple command-line interpreter for simplicity. Finally, the command cache holds the last executed commands and other responses from controlled devices. This cache allows multiple compromised smart controllers to fetch the same execution message stored in the command server. Additionally, the command cache allows Mallory to disconnect while an attack is active to retrieve the responses from an attack at a later time.

\textbf{\modulethreetitle.} The \modulethree serves as the attack vector and performs the bulk of malicious operations in \name. When the \modulethree is active, the driver contacts the \moduletwo for new attacks to execute, multiple controlled devices may contact the same server. The \modulethree module contains two components: the \textit{driver logic} and the \textit{malicious payload}. The driver logic can be seen as a standard operating code to allow the malicious driver to appear and operate as a non-malicious driver. A malicious driver must appear as if it is benign, providing all standard operations a legitimate driver offers. The malicious payload component contains all the code required to execute attacks. The malicious payload may cause memory leaks, perform malicious requests to servers, eavesdrop, and otherwise execute any operation beneficial to Mallory.

\textbf{\modulefourtitle.} Serving as the host of malicious drivers, the \modulefour is the EIoT system itself. These include any system in a smart building, luxury home, or office, which may be compromised by a malicious driver. The \modulefour is connected to the Internet and contains all of the devices of the affected smart system, including the centralized controller. As part of the \name's end-to-end implementation, the \modulefour is one of the targets compromised by the \moduleone during attack activation.

% --------------------- IMPLEMENTATION ----------------------------

\section{Evaluation and Realization of \name Attacks}
\label{sec:Implementation}

\begin{figure}[t]
\centering{\includegraphics[width=0.47 \textwidth]{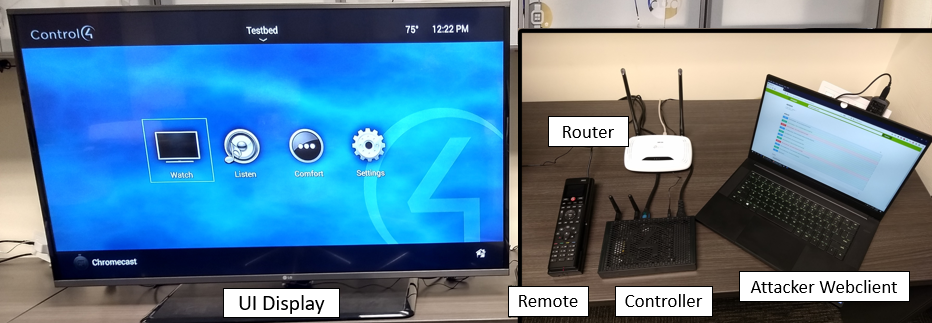}}
    \caption{EIoT system testbed used to implement \name attacks in a smart building setting.}
\label{fig:testbed}
\vspace{-0.2in}
\end{figure}

\begin{table}[b]
    \centering
    \caption{Hardware \& software used in \name attacks implementation and testing.}
    \scalebox{0.90}{
    \begin{tabular}{cc}
    \hline
       Hardware & Software \\ 
       \hline
       Control4 EA-1 Controller & Microsoft Visual Studio Code 1.4.11  \\
       \rowcolor{Gray}
       Control4 SR-260 & Control4 Driver Editor 3.0.1 \\
       LG 49LX570H & Control4 Composer 2.10.6 \\
       \rowcolor{Gray}
       TP-Link TL-WR841N Router & Jersey JAX-RS with Swagger.io\\
       Razer Blade 15 Laptop & Amazon AWS Elastic Beanstalk\\
    \hline
    \end{tabular}}
    \label{tbl:testbed&software}
\end{table}

In this section, we demonstrate the implementation of \name attacks on our realistic EIoT testbed. Further, we evaluated the effects of \name attacks on the EIoT system in detail. 

\subsection{\name Implementation on Real EIoT Devices}
\begin{figure}[t]
\centering{\includegraphics[width=0.45 \textwidth]{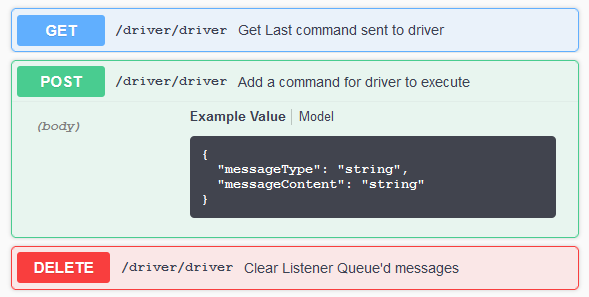}}
    \caption{Swagger interface for \name remote attack execution with JSON object. The \textit{messageType} field determines the attack type and \textit{messageContent} for extra parameters.}
\label{fig:webserver}
\vspace{-0.2in}
\end{figure}

We created a malicious television EIoT driver (as detailed in Figure \ref{fig:architecture}) and an EIoT system testbed with real Control4 devices as shown in Figure \ref{fig:testbed}. Control4 was selected as it is one of the most popular EIoT systems available in the market, named a leading brand in EIoT for five years in a row \cite{control4leading}. The testbed included vendor-specific devices and is configured to function as a small EIoT system (Table \ref{tbl:testbed&software}). We utilized Driver Editor, a tool available for the development of drivers in Control4 \cite{drivereditor}. Additionally, we used LUA, an open-source programming language which is the core development platform of Control4 drivers \cite{driverlua}. We configured the EIoT system using Control4's Composer 2.10.6 and with an EA-1 as the main controller. To grant Internet access to the devices included in the testbed, we configured a network with the TP-Link TL-WR841N Router. We verified the running version of LUA in Control4 devices as LUA 5.1 programmatically (executing a script which returned the running LUA version). To implement \name realistically, we created a command and control webserver with a RESTful API in JAX-RS hosted in Amazon AWS. The Swagger-based web interface for the server can be seen in Figure \ref{fig:webserver}

\textbf{Execution JSON Object. } A JSON object is used by \name modules to exchange attack details. The JSON response object consists of two fields. The \textit{messageType} field is the attack type to be executed. The \textit{messageContent} field contains additional information such as the target URL to attack.

\textbf{Command and Control Webserver.} To implement \name realistically, we created a webserver that could be queried by the malicious driver for attack commands. The server hosts a RESTful API implemented using JAX-RS and Swagger add-on as a UI interface. The Swagger-based web interface for the server can be seen in Figure \ref{fig:webserver}. Primarily, the webserver has one endpoint \textsc{/driver/driver} with POST, GET, and DELETE. In our implementation, the REST request types were used as follows:

 \begin{itemize}
     \item \textit{GET:} Fetches the last JSON object received by the API, and is used by the driver to poll for the last command received.
     \item \textit{POST:} Submits a new JSON object to be stored by the API, overwriting the previous values.
     \item \textit{DELETE: } Clears the stored object fields, setting both the \textit{messageType} and \textit{messageContent} to \textsc{Null}.
 \end{itemize}

\subsection{Software Modules}

To execute \name attacks and implement the malicious driver, we created a number of LUA software modules.

\textit{1) Remote Polling Module: } The remote polling module awaits commands from an attacker-managed server which issues commands to execute specific \name-based attacks. As with all software modules, the remote polling module was written in LUA. Pseudocode to demonstrate the polling and selection process can be seen in Algorithm \ref{alg:polling}.  As a traditional REST client, the first initialization request is ``DELETE: [URL]/driver/driver'' which clears the command cache in the webserver (Line 2). Once initialized, the module executes as a loop every three seconds. The command server's address is polled with the request ``GET: [URL]/driver/driver'' and the response JSON object stored into a local cache (Line 4). This newly received command is compared to the last command received if the command is different (Line 5), then the attack specified is initiated (Line 6). After the execution, the local cache and the server JSON messages are cleared (Line 7). After the execution is finished, the loop waits and initiates again.

\newcommand{\LineComment}[1]{\hfill // \textit{#1}}

\textit{2) LUA Hashing Module: } A notable challenge for the development of \name was the creation of a hashing module to perform cryptographic hashing (SHA-256) operations in a LUA-based driver. LUA 5.1 does not support the bitwise operations from the standard libraries; thus, the hashing algorithm had to be adapted for this version of LUA. 
We utilized several sources of code for the implementation of SHA-256, commonly used for password hashing and cryptocurrency mining \cite{bitcoinmining}. The SHA-256 hashing algorithm was implemented in pure LUA for this module to effectively test cryptography-based threats in \name.

\textit{3) Memory Exhaustion Module: } To perform attacks, we created a software module that would allow an attacker to expend system resources in the controller. This module was implemented as a LUA table data type and a loop that iterated over itself, adding content to the table to expend system resources. This code caused a DoS condition in the target system.

\textit{4) Network Request Module: } The network request module was created so that \name could perform GET requests to remote URLs. The module was implemented using Control4 specific API command \textsc{C4:urlGet()} to execute a GET request to a given endpoint. The request is placed in a loop, in effect, this allows the attacker to perform a set number of requests or continue making requests indefinitely. While the API command returns the fetched data, the data is only used for confirmation of a successful query.

\begin{algorithm}[t]
\scriptsize
    \caption{\name attack polling algorithm}
    \label{alg:polling}
    \begin{algorithmic}[1]
            \State Initialization; \LineComment{Initializes driver variables}
            \State DELETE: [URL]; \LineComment{Clear server cache}
            \While{true} \LineComment{Operation loop}
                \State LocalCache <- GET: [URL]; \LineComment{Get server cache }
                \If{New Command in LocalCache}
                    \State Execute Attack Specified;
                    \State DELETE: [URL]; \LineComment{Clear server cache} 
                \EndIf
                \State Wait 3 Seconds;
            \EndWhile
    \end{algorithmic}
\end{algorithm}

\subsection{\name-based Attacks}

In this sub-section, we realize the \name attacks and discuss the results and implications of each attack. All of the attacks presented begins with the initiation of the driver and by polling the server. The attacks were executed using an EA-1 controller with a malicious driver querying the AWS-hosted server. The Razer Blade 15 laptop was used for the remote execution of the attacks. 

\textbf{Attack 1: Denial-of-Service.} 
This attack was developed to demonstrate that Threat 1 is possible through \name. This attack implements a DoS condition in a local system by causing memory exhaustion in the host controller.

\textit{Step 1 - Activation.} The activation of the driver was executed remotely through the attacker's web interface. With this web interface, the \textsc{messageType} field of the JSON object was set to ``DOS'' to initiate a local DoS attack.

\textit{Step 2 - DoS Payload.} As the driver polls the server with the remote polling module, the activation message was successfully interpreted by the driver, and the attack was initiated. The action activated the memory exhaustion module and begins to consume system resources in the target device.

\begin{figure}[t]
\centering{\includegraphics[width=0.45 \textwidth]{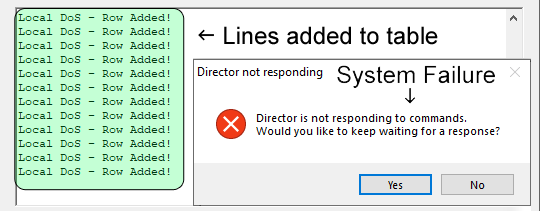}}
    \caption{Attack 1 (Memory Exhaustion) implementation results. Figure shows items being inserted into a LUA table, creating resource exhaustion.}
\label{fig:localdos}
\vspace{-0.25in}
\end{figure}

\textit{Evaluation:} This attack was entirely successful as the device hosting the driver (the controller) was rendered inoperable within five seconds of activation, affecting the controller in two ways. First, any configuration software connected to the main controller lost connection and was locked up. Figure \ref{fig:localdos} shows the configuration software losing connection with the controller during our attack. Second, any communication with the central controller was interrupted, meaning that a user would have no way to use the EIoT system once this attack was active. On-screen interfaces (e.g., Television UI interface, computer interface) and handheld remotes lost communication with the main controller, preventing access to any of the other smart devices integrated with the controller. With no form of disabling the loop, the only option to re-establish the device was to power cycle. It is even possible to run this attack on the controller's initiation, effectively rendering the device inoperable even after rebooting. 

\begin{figure}[t]
\centering{\includegraphics[width=0.48 \textwidth]{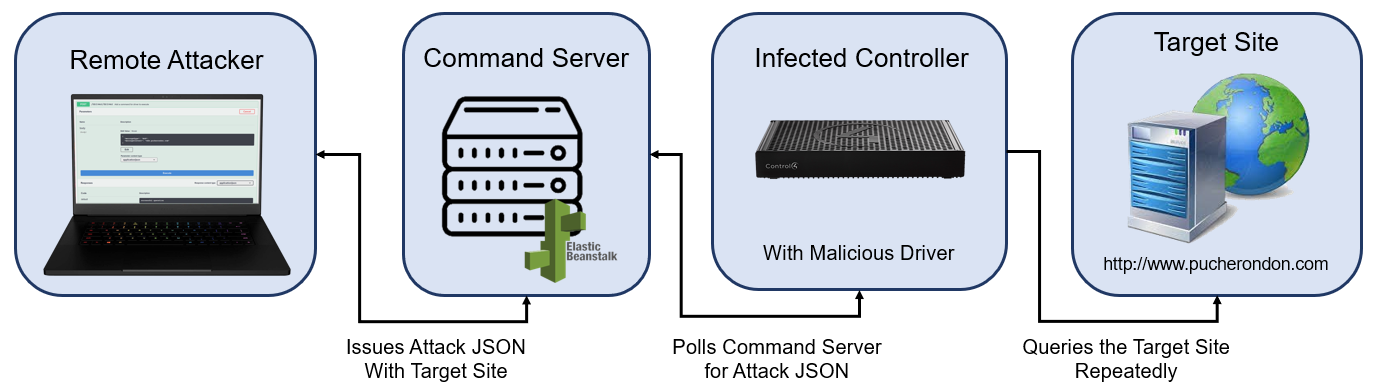}}
    \caption{Implemented botnet attack model for Attack 2. The remote attacker initiates the attack as shown in this figure.}
\label{fig:botnet}
\vspace{-0.2in}
\end{figure}

\textbf{Attack 2: Remote Control.} The remote control attack serves as a way to demonstrate the feasibility of Threat 2. Primarily, we show that a remote attacker (Mallory) can take control of one (or many) devices and command them to make continuous malicious requests to a specific server, negatively impacting the critical servers. We follow Figure \ref{fig:botnet} for the following steps.  

\textit{Step 1 - Activation.} This attack was activated through the use of the available web interface by the remote attacker laptop. The JSON object \textsc{messageType} field was set to ``BOT'' and \textsc{messageContent} to ``www.pucherondon.com'' to initiate a repeated querying to the target site. 

\textit{Step 2 - Execution.} As we did not want to disrupt the functionality of the target webserver, we used the network request module with a loop of ten requests to the target webserver. Once the JSON object was received, the requests were made to an external website ``www.pucherondon.com''. All of the requests were successful on the target webserver.

\textit{Evaluation.} We evaluate this attack by the success of remote attack activation and the requests to the target site. The attack received the remote commands from the remote attacker laptop and then performed web requests upon the target website without any issues. Thus, the commands issued by the remote attacker were executed on a target site. While the request loop was kept to ten executions, one can easily increase to any number of requests. The purpose of this test was to demonstrate that remote activation and querying of a page is possible via \name attacks. Additionally, if there are multiple controllers with the malicious driver, infected devices could perform a more effective distributed attack on a target webpage by increasing the number of requests, creating a Distributed Denial-of-Service (DDoS) attack. The goal of this implementation is to demonstrate the remote attacks are possible, which was proven by our attacks. Additionally, scaling is very straightforward, which can be done by increasing the number of compromised controllers with malicious drivers available to the attacker. It is possible that complex EIoT deployments may be compromised with drivers and used for DDoS attacks in a similar manner to Mirai with EIoT systems if not properly secured \cite{mirai}.

\textbf{Attack 3: Malicious Resource Farming.} Resource farming attack was developed to demonstrate that system resources may be used to Mallory's benefit for a purpose such as bitcoin mining. Currently, bitcoin uses a double hash SHA256 operation (Equation \ref{equ:mining}) where, \textit{B} represents recent transactions, \textit{N} represents a nonce, and \textit{T} is the target value \cite{bitcoinmining}. 

\vspace{-0.1in}

\begin{equation}
\label{equ:mining}
T > SHA256(SHA256(B.N))
\end{equation}

For \name, we performed the required cryptographic operations used in bitcoin mining. To demonstrate that such operations can be done within a driver, we executed multiple hashing operations in the infected device.

\textit{Step 1 - Activation: } This attack was activated similar to the previously introduced attacks, using the web interface and a JSON object request. To initiate this attack, the \textsc{messageType} field was set to ``MIN'' in the outgoing JSON object from the client computer, the driver interpreting the change as a request to perform mining-based cryptographic operations.

\textit{Step 2 - Execution:} After the internal remote polling module processed by the driver, the \textsc{messageType} field, the driver calls the LUA hashing module which executes ten hashing operations using controller resources. Similarly to cryptocurrency mining, the hashing operations were performed with a static \textit{B} value and random nonce values for \textit{T} in each iteration.

\begin{figure}[t]
\centering{\includegraphics[width=0.45 \textwidth]{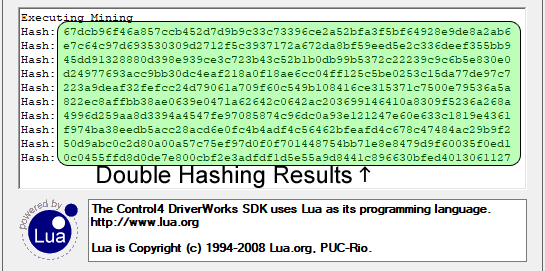}}
    \caption{Hashing process as executed by \name attacks.}
\label{fig:mining}
\vspace{-0.2in}
\end{figure}

\textit{Evaluation:} The driver managed to perform all hashing operations successfully. Figure \ref{fig:mining} shows a sample of ten mining operations executed in the malicious driver. In the case of multiple devices infected with malicious drivers, the number of machines performing hashing operations on the attacker's behalf could be increased. This type of resource farming attack could negatively impact the performance of an EIoT system depending on how many cryptographic operations are executed per minute. The number of hashing operations per minute can also be adjusted to avoid detection.

\textbf{Discussion and Findings}
As \name attacks were developed and tested, we demonstrate that insecure software development, lack of built-in security, and untrusted drivers result in malicious activities. We have found drivers to be a viable threat vector; thus, we have coordinated and shared of our findings with Control4 for further discussion. Without any form of verification, an integrator may download a compromised driver and allow a malicious actor to compromise an EIoT system to her benefit. All of the proposed attacks were implemented successfully, the implications which could negatively impact EIoT systems. In Attack 1, we demonstrate that an entire system can be rendered unusable at the command of an attacker and is possible due to the ability of drivers to consume system resources without limitations. The attack presents a viable method of disabling access to security systems, gates, doors, or any other system which is integrated into an EIoT system. For instance, if gate access or panic button is controlled purely through an EIoT system, a user will not be able to operate the gate access or a panic button while a DoS attack is active. Attack 2 is made possible due to the lack of limitations on connections to external websites and shows how an attacker can perform DDoS-type of attacks on target webpages using multiple controllers.

There have been documented cases of malware purposely accessing illegal websites to frame the system owners \cite{malwareframe}. An attacker with a compromised EIoT system may request illegal websites and frame the owners for illicit activity. In this work, we cited an example of one use-case of cryptographic operations as cryptocurrency mining. These results also imply that an attacker may also perform any other hardware-intensive actions such as password cracking. Ultimately, our implementations show that drivers as attack vectors have many possibilities. Attack 3,  is possible due to a lack of restrictions in the LUA implementation and unfettered access to system resources. Further, with processor-intensive operations, a compromised controller could also be used for cracking hashed passwords. An attacker with a list of passwords to crack could use the processing power of compromised controllers to attempt to reverse password hashes, a very similar operation to cryptocurrency mining. As \name-style attacks present a substantial negative impact on EIoT systems, acknowledging these threats and finding solutions should be of utmost importance. We believe that a security verification mechanism is needed for EIoT systems that verify the integrity and origin of the drivers. In addition, an EIoT system controller needs inherent security mechanisms that limit external communications, resource consumption, and access to high-end system resources.

% --------------------- DISCUSSION ----------------------------

\section{Discussion}
\label{sec:Discussion}
In this section, we discuss the implications of these attacks and possible defense mechanisms for \name attacks. 

\textbf{\name-based Attacks.} With \name, we explored possibilities of attacks that could be implemented through EIoT smart device drivers. Depending on the capabilities of the driver, in addition to the attacks demonstrated in this work, it is possible for a driver to act as a keylogger, capturing key-presses relayed to a device from any interface. For instance, if a user has a media device with login credentials for web services (such as Netflix in an integrated AppleTV) an attacker may be able to capture those credentials. Specifically, if a user uses an infected driver to communicate to the media device and enter their password with the arrow keys on an on-screen keyboard, a malicious driver could intercept the key-presses and capture the user's credentials. Another possible attack, depending on the driver implementation, may involve weak script interpreter implementations. If there are weaknesses to the interpreter, an attacker may be able to perform injections through a driver using known vulnerabilities. 

\textbf{Challenges in Standardization.} 
One of the biggest challenges in EIoT systems and IoT as a whole is standardization. There are countless companies, protocols, and implementations of many technologies depending on the vendor. Drivers are no different;  how drivers are implemented from system to system are different. As attackers become more sophisticated, manufacturers cannot rely on a closed-source system for security. However, having multiple vendors agree in a standard to interface with devices is not an easy task. Efforts to standardize control in protocols such as IP, such as Project Connected Home are ongoing, and projects such as these do not address EIoT system drivers \cite{connectedHome}. An effort to standardize how drivers and how they are implemented would be the first step towards security. Further, with many EIoT systems deployed in the world, legacy systems present a problem to developing defense mechanisms against any new threats. By definition, many of these legacy systems can not be upgraded to the latest security practices \cite{legacysystem}. A great number of systems may not longer be supported or their vendor is no longer in business (e.g., Litetouch, X10-technology) \cite{LitetouchDead}. As such, there are many legacy EIoT systems which may be too costly or too impractical to upgrade. Legacy EIoT raises the issue that these systems cannot be patched easily and be compromised by a knowledgeable attacker. Defense mechanisms must consider the limitations that come with legacy systems and how to secure them.

\textbf{Risk Awareness.} Most vendors have documented best practices for the installation of their devices, discouraging risky configurations such as port forwarding directly to their controllers. However, installers will still port-forward their devices, exposing them to the Internet as it is the easiest solution. As many controllers were not designed to be connected directly to the Internet, they could become compromised by an attacker if exposed. First, the usage of VPNs needs to be documented in a proper manner for remote access to devices. Second,  EIoT system integrators should be wary of drivers on the Internet and favor trusted drivers provided by EIoT system vendors. We hope this work besides motivating further research work on protecting EIoT systems from novel types of attacks, can raise awareness for integrators on what malicious code is able to do, and allow them to evaluate the risks of using unverified drivers. Second, integrators must be aware that because one version of a driver is verified, updated versions may not. This could create a false sense of security, as an attacker may be able to verify a benign driver, then link to their own page for an updated, malicious version of a driver.  

\textbf{Comprehensive Driver Validation.} As driver development for every vendor is different, vendor certification of drivers is the most effective step towards the security of the EIoT system drivers. As of now, the development and distribution of unverified drivers come without any form of source control, standards, or code analysis. Existing driver certification needs to evaluate beyond functionality and consider that code could be implemented maliciously. Additionally, EIoT system vendors could allow for the submission of drivers and perform code analysis to drivers submitted to their platform. Such an idea would create a larger number of drivers available to vendors. Vendors should then highlight that unverified drivers should be used at the integrator's own risk.

% --------------------- RELATED WORK ----------------------------

\section{Related Work}
\label{sec:RelatedWork}

Research and active attacks against smart devices have been an ongoing topic of research in recent years. Work by Zhang et al., presents an overview of common vulnerabilities in IoT such as weak authentication, over-privilege and implementation flaws in connected devices \cite{ZhangIoT}. Within the scope of smart homes, work by Abrishamchi et al., summarizes side-channel threats in smart home systems~\cite{smarthomesidechannel}. Work presented by Koh et al. highlights how smart building applications are often over-privileged for their purpose. Defense mechanisms have been proposed in response to threats in smart buildings, homes and appliances \cite{rimor, trustweatherdata, smartlighting, berkaymagazine}. As early as 2013, some works highlight various threats in smart devices and note that attackers are in constant search of new methods to infect smart devices with malware or to detect them \cite{smarthomemalware, saint, daint, lopez2017survey, convolution, smartgridjournal, patent, patent2, aegis, kratos, madiot, ziot, icc, usbjournal, iotdots, heka}. Additionally, research in alternative attack vectors such as HDMI and USB exists, which can provide attackers with an expanded attack vector as CEC drivers gain more popularity \cite{hdmiwalk,denney2019usb,HDMIWatch}. 

The rise of Internet-connected devices has given malicious actors an unprecedented number of machines to overtake with botnet malware. One of the greatest examples is the Mirai botnet, which crippled multiple sites during its peak \cite{miraiimpacts}. Research on the technical aspects and impact of this botnet revealed this was possible due to insecure practices and vulnerabilities in exposed devices \cite{mirai}. Further, botnets have become a lucrative business for attackers using traditional malware \cite{botnets}. Additionally, to attack botnets, mining botnets have gained popularity with the rising use of cryptocurrency. Case studies by the Cyber Defense Group explain how malicious attackers use malware-compromised machines to mine cryptocurrency at user's expense \cite{mining}.

\noindent\textit{Our work differs from the previously discussed works as \name focuses on the insecurity of EIoT system drivers, an attack vector which has been largely unexplored. In contrast to injecting malicious code into an operating system, our attacks rely entirely on weaknesses available through EIoT system design and lack of secure development practices. We focus on the exploitation of drivers to a remote attacker's benefit and create proof-of-concept implementations of a malicious driver. Specifically, we present three specific threats that are possible to implement with a malicious driver: (1) DoS attacks on the host system, (2) remote control of a target EIoT system, and (3) the malicious farming of system resources for unauthorized activities (e.g., bitcoin mining). With \name, we demonstrate it is possible for an attacker to assume control of the EIoT system in a malicious nature, solely through the use of drivers.}

% --------------------- CONCLUSIONS ----------------------------

\vspace{-0.05in}
\section{Conclusion}
\label{sec:Conclusions}

Recent years have seen a dramatic rise in IoT systems and applications that enabled billions of commodity IoT devices to empower smarter settings in buildings, offices, and homes. Although commodity IoT devices are employed by ordinary end-users in small-scale environments, more reliable, complex, customized, and robust solutions are required for enterprise customers. Those solutions called EIoT are offered by dedicated vendors. With the higher price, customization, robustness, and scalability of EIoT systems, they are commonly found in settings such as smart buildings, government or private smart offices, academic conference rooms, luxury smart homes, and hospitality applications. As EIoT systems require specialized training, software, and equipment to deploy, many of these systems are closed-source and proprietary in nature. This has led to very little research investigating the security of EIoT systems and their components. In effect, EIoT systems in professional smart settings (e.g., smart buildings) present an unprecedented and unexplored threat vector for an attacker. In this work, we explored EIoT system vulnerabilities and insecure development practices, specifically, the usage of drivers as an attack mechanism. We implemented an EIoT system testbed in a smart building setting and introduced \name, a novel attack mechanism to show that it is possible for a malicious actor to easily attack and command EIoT system controllers using malicious drivers. Specifically, with \name, an attacker may cause DoS conditions, take control of EIoT system controllers, and remotely abuse the resources of the such systems for illegal activities (e.g., bitcoin mining, etc.). With this work, we raise awareness on the (in)secure development of the drivers that control EIoT systems, the consequences of which can largely impact EIoT systems and targeted smart buildings as a result. Additionally, we discussed the (in)security of these drivers, security implications, and possible counter-measures. To the best of our knowledge, this is the first work to systematically analyze the (in)security of the EIoT deployments and clearly demonstrate the vulnerabilities in this ecosystem. 

\section*{Acknowledgments}
This work is partially supported by the US National Science Foundation (Awards: NSF-CAREER-CNS-1453647, NSF-1663051). The views are those of the authors only.

\bibliographystyle{ACM-Reference-Format}
\bibliography{references}

\end{document}